**Observation of Self-Cavitating Envelope Dispersive Shock Waves in Yttrium Iron Garnet Thin Films**

P. A. Praveen Janantha,[1] Patrick Sprenger,[2] Mark A. Hoefer,[2] and Mingzhong Wu[1*]

[1]*Department of Physics, Colorado State University, Fort Collins, CO 80523, USA*

[2]*Department of Applied Mathematics, University of Colorado, Boulder, CO 80309, USA*

The formation and properties of envelope dispersive shock wave (DSW) excitations from repulsive nonlinear waves in a magnetic film are studied. Experiments involve the excitation of a spin-wave step pulse in a low-loss magnetic $Y_3Fe_5O_{12}$ thin film strip, in which the spin-wave amplitude increases rapidly, realizing the canonical Riemann problem of shock theory. Under certain conditions, the envelope of the spin-wave pulse evolves into a DSW that consists of an expanding train of nonlinear oscillations with amplitudes increasing from front to back, terminated by a black soliton. The onset of DSW self-cavitation, indicated by a point of zero power and a concomitant 180° phase jump, is observed for sufficiently large steps, indicative of the bidirectional dispersive hydrodynamic nature of the DSW. The experimental observations are interpreted with theory and simulations of the nonlinear Schrödinger equation.

In a nonlinear, hydrodynamic medium where dispersion dominates over dissipation, an initial, abrupt increase in a physical quantity (e.g., the water height) can evolve into an expanding waveform composed of a soliton edge followed by diminishing amplitude modulations. Such a nonlinear wavetrain is called a dispersive shock wave (DSW).[1] The phenomenon of DSWs is ubiquitous in nature, appearing in dispersive media as diverse as the ocean,[2,3] intense laser light,[4,5,6] electron beams,[7] ultra-cold atoms,[8,9,10] and viscous fluids.[11] It is the superfluid or dispersive hydrodynamic analog of a viscous shock wave in a gas.[1] In contrast to the entropy production and energy dissipation due to friction in viscous shock waves, however, a DSW conserves energy, converting the potential energy of an initial jump into the kinetic energy of nonlinear oscillations with no entropy generation.

The simplest dispersive hydrodynamics are unidirectional or bidirectional; bidirectional waves correspond to the existence of a fluid velocity in addition to a fluid density and the co-existence of two distinct wave families with different velocities. A universal model of bidirectional dispersive waves is the nonlinear Schrödinger (NLS) equation,[1] which can effectively reproduce, for example, the wavefunction dynamics of a Bose-Einstein condensate[8,9,10] and the envelope dynamics of light beams in nonlinear Kerr media.[4,5,6] Celebrated features of the NLS model include bright solitons and modulational instability in attractive media[12] and dark solitons and DSWs in repulsive media.[13,14] A canonical, textbook problem in shock wave theory is the Riemann problem, consisting of an initial, steep change in the hydrodynamic medium's thermodynamic variables, e.g., density and velocity.[15] For the repulsive NLS equation, the Riemann problem can result in the generation of an envelope DSW with a dark soliton edge.[1,13,14] Further, the derivative profile of the DSW phase oscillates in tandem with the amplitude. In terms of dispersive hydrodynamics, the square of the DSW amplitude corresponds to a fluid density, while the phase gradient is analogous to a fluid velocity. A notable property of sufficiently large NLS DSWs is the generation of a zero density (or a vacuum) point with a distinct 180° phase jump signature. This corresponds to the spontaneous cavitation of a DSW, a property unique to dispersive, vis-à-vis viscous shock waves. What's more, in the reference frame of the dark soliton edge, a cavitating DSW exhibits a surprising feature: the upstream and downstream velocity fields point into the DSW from both sides.[1,14] One can see that the phase plays a rather major role in bidirectional dispersive hydrodynamics in general and self-cavitating DSWs in particular.

Experimentally, DSWs have been observed in various bidirectional media.[4,5,6,8,9,10,16,17] All observations to date, however, have been limited to the evolution of the amplitude (or density) of either localized or periodic pulses, while the phase (or velocity) features have never been studied. In other words, although previous experiments have seen oscillations that appear to go to zero amplitude, they did not measure the phase to make a definitive determination of the self-cavitating signature of DSWs.

This letter reports the first observation of *bona fide* self-cavitating envelope DSWs. The experiments use surface spin waves in a magnetic $Y_3Fe_5O_{12}$ (YIG) thin film strip that exhibits repulsive nonlinearity and low damping,[18] enabling bidirectional dispersive hydrodynamics. The experiments demonstrate envelope DSWs resulting from a step amplitude increase, realizing the dispersive hydrodynamic equivalent of the shock tube problem of viscous gas dynamics. Self-cavitation is characterized through both direct amplitude and phase measurements as well as simulations. The results not only advance the fundamental understanding of DSW physics, but also help interpret various hydrodynamic effects, such as turbulence and de-coherence, in a variety of bidirectional nonlinear wave systems.[19,20,21]

Figure 1(a) shows a schematic of the experimental setup. The setup includes a 36.5-mm-long, 1.3-mm-wide YIG thin film strip cut from a YIG wafer grown on $Ga_3Gd_5O_{12}$. Two 50-μm-wide, 2-mm-long microstrip transducers are placed above the YIG strip for the excitation and detection of spin waves, and their separation ($l$) is 20.8 mm. The YIG film strip is magnetized to saturation by a magnetic field of 1323 Oe, which is in the YIG plane and perpendicular to the YIG strip length. This film/field configuration supports the propagation of surface spin waves[22,23,24] with repulsive nonlinearity.[18] For the DSW measurements, the excitation transducer is fed with a microwave step pulse whose power is $P_1$ just before the step and $P_2$ just after, as indicated in Fig. 1(a). Such a microwave pulse excites a spin-wave step pulse in the YIG film. During the measurements, $P_1$ and $P_2$ are varied over 1 μW–80 mW, but the carrier frequency $f_0$ is fixed to 6.045 GHz. The output signals are measured with an oscilloscope.

Figures 1(b), 1(c), and 1(d) show the characteristics of the YIG device in Fig. 1(a). Figure 1(b) presents the amplitude of the S-parameter $\mathbf{S}_{21}$ measured at a microwave power of 20 μW over a frequency ($f$) range of 5.75-6.30 GHz. In Fig. 1(c), the blue curve presents the spin-wave dispersion curve determined from the phase of $\mathbf{S}_{21}(f)$, namely, $\phi(f)$, while the red curve shows a theoretical fit to the dispersion equation. To obtain the experimental curve, the spin-wave wavenumber $k(f)$ was calculated from $\phi(f)$ using the relation $\phi(f)=k(f)l+\phi_0$ ($\phi_0$, a phase constant) and taking $k$=0 at the low cut-off frequency $f_{cut}$=5.792 GHz of the transmission. The fitting used[18,23]

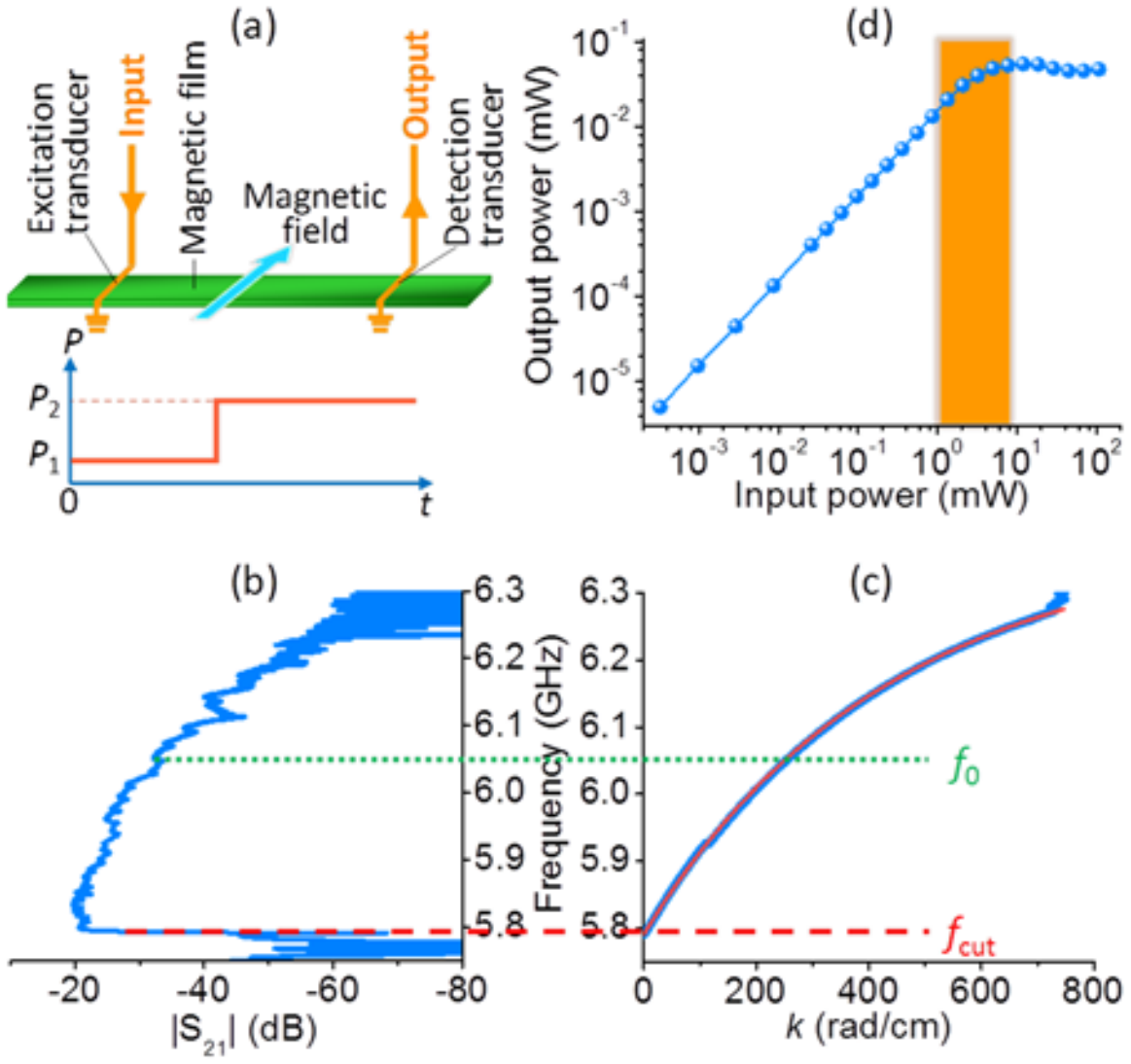

**Fig. 1. Experimental configuration and spin-wave characteristics.** (a) Schematic of the experimental setup. (b) Transmission response of the YIG film strip measured at an input power of *P*=20 μW. (c) Experimental (blue) and theoretical (red) dispersion curves of spin waves in the YIG film strip. (d) Output power of the YIG film strip device measured as a function of *P* at 6.045 GHz.

$$2\pi f = |\gamma|\sqrt{H_0\left(H_0 + 4\pi M_s\right) + \left(\tfrac{1}{2}4\pi M_s\right)^2\left(1 - e^{-2kd}\right)} \quad (1)$$

where $|\gamma|$ is the absolute gyromagnetic ratio, $H_0$ is the field, $4\pi M_s$ is the saturation induction of the YIG, and $d$ is the YIG film thickness. The horizontal dotted and dashed lines in Figs. 1(b) and 1(c) indicate $f_0$ and $f_{cut}$, respectively. Figure 1(d) gives the output power as a function of the input power measured at $f_0$. The shaded area indicates the $P_2$ range in which pronounced DSW excitations were observed.

The data in Fig. 1(b) indicate a spin-wave passband from 5.79 GHz to ~6.20 GHz, in which $f_0$ is centrally located. The transmission profile is relatively smooth, indicating that the spins on the YIG film surfaces are unpinned and a repulsive nonlinearity is expected for the entire frequency range. In films with pinned surface spins, the nonlinearity is repulsive only in narrow frequency ranges.[25,26] The fit in Fig. 1(c) is nearly perfect and yields $|\gamma|$=2.88 MHz/Oe, $4\pi M_s$=1870 G, and $d$=11.0 μm. The $4\pi M_s$ value is slightly larger than the bulk value (1750 G), mainly due to the assumption of a zero anisotropy field in Eq. (1). The dispersive characteristics of a wave are determined by the dispersion coefficient $D = \frac{\partial^2(2\pi f)}{\partial k^2}$. Using the experimental curve in Fig. 1(c), one obtains $D$=-11.8×10$^3$ cm$^2$/(rad·s) at $f_0$=6.045 GHz, very close to the theoretically calculated value -11.3×10$^3$ cm$^2$/(rad·s). The data in Fig. 1(d) indicate that the spin wave is linearly damped in the 10$^{-4}$-1 mW input power range but is nonlinearly damped for powers larger than 10 mW. The nonlinearity here derives mainly from four-wave interactions.[12,27]

The DSW data are presented in Figs. 2-5. Figure 2 depicts the main results. Figures 2(a) and 2(b) present the input and output signals, respectively. In each column, the top and bottom panels present the amplitude and phase profiles, respectively. The phase profiles are relative to that of a continuous wave with frequency $f_0$=6.045 GHz. In Fig. 2(c), the top panel shows the square of the amplitude data in Fig. 2(b), and the bottom shows the same phase data in Fig. 2(b). The red curve in Fig. 2(c) shows a fit to the square of a black soliton profile[12]

$$u(t) = u_0 \sqrt{1 - \mathrm{sech}^2\left[ u_0 v_g \sqrt{\frac{N}{D}} (t - t_0) \right]} \tag{2}$$

where $u$ denotes the spin-wave amplitude, $u_0$ is the black soliton background amplitude, $v_g = \partial(2\pi f)/\partial k$ is the group velocity, $N = \partial(2\pi f)/\partial(|u|^2)$ is the nonlinearity coefficient,[18] and $t_0$ is a time constant. Calculations using the experimental parameters yielded $v_g = 5.0\times10^6$ cm/s and $N = -2.9\times10^9$ rad/s. $u_0$ and $t_0$ are two fitting parameters, equal to $3.28\times10^{-2}$ and 277 ns, respectively, for the fit in Fig. 2(c). The vertical dashed line in Fig. 2(c) indicates the location of the black soliton center. All the data were taken with $P_1$=0.14 mW and $P_2$=3.47 mW.

The data in Fig. 2 indicate that an envelope DSW is formed consisting of a train of dark soliton-like dips, with broadening widths from small-amplitude oscillations to termination at approximately a black soliton. The observed soliton has almost zero amplitude at its center, a profile that can be fitted with the black soliton function, and a phase jump of 175° that is close to the theoretically expected 180°. All of these features are signatures of black solitons. The fact that the DSW is terminated by a black soliton, not a gray soliton, indicates self-cavitation and the formation of a vacuum point where the phase gradient or fluid velocity is theoretically infinite.[14] In the insert, the derivative of the phase profile shows an oscillation behavior, with both the oscillation period and amplitude increasing from the front to the back. Note that the derivative at the soliton center is beyond the vertical scale and the oscillation reverses polarity, indicative of a vacuum point. Following the DSW is a rarefaction wave (RW), a smooth, expansion wave exhibiting weak oscillations in both amplitude and phase.[8] The long-time evolution of an initial step in the amplitude for the NLS model is analogous to the shock tube problem of gas dynamics or the dam break problem of hydrodynamics and has been classified,[14] resulting in a DSW connected to a RW by an intermediate state. According to the classification and the experimental parameters in this work, one can expect a faster DSW followed by a slower RW. This prediction is evident in Fig. 2. The phase in Fig. 2(b) reveals an approximately constant, negative slope for the RW. This frequency shift is due to nonlinear dispersion and corresponds to a concomitant positive *k* shift, as expected by the initial spin-wave step that maintains a constant frequency.

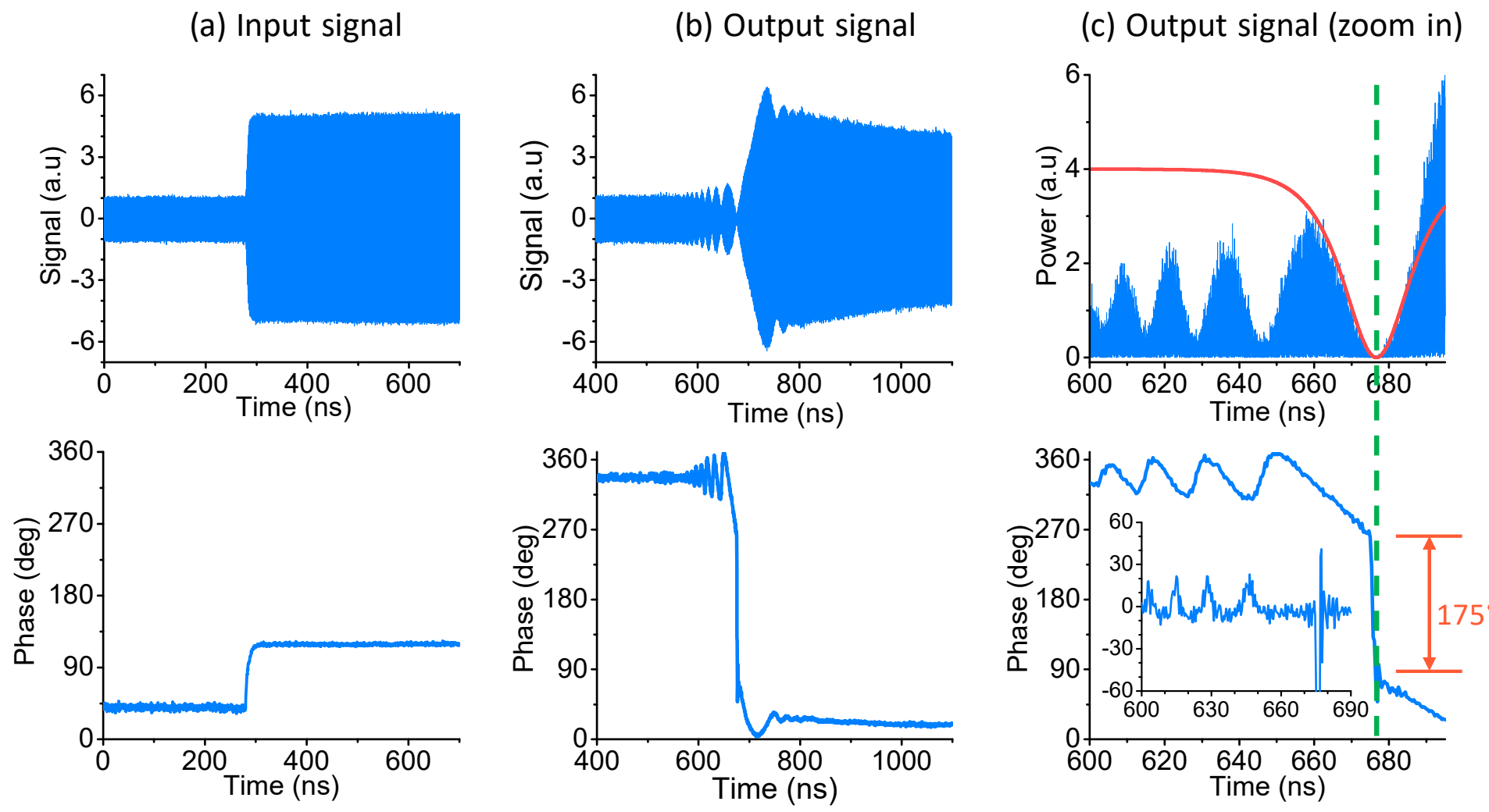

**Fig. 2. Demonstration of a cavitating envelope DSW.** (a) Input signal: top – amplitude; bottom - phase. (b) Output signal: top – amplitude; bottom - phase. (c) Zoom-in display of the output signal shown in (b): top – power (amplitude square); middle – phase; insert – phase derivative (degree/ns). The red curve in (c) shows a numerical fit to a black soliton profile.

Figures 3 and 4 present data revealing that DSW formation is sensitive to the characteristics of the initial step. Figure 3 presents data measured with $P_2$=3.47 mW while the ratio $P_2/P_1$ ranges from ~1.6 to 50. As $P_2/P_1$ is increased, the soliton dip at the DSW edge deepens, with a gray soliton for $P_2/P_1$ up to 7.6. A black soliton edge and the corresponding vacuum point appear for $P_2/P_1$=16, which is consistent with theory that predicts self-cavitation for $P_2/P_1$>9.[1,14] Higher $P_2/P_1$ leads to the migration of the vacuum point away from the soliton edge and into the DSW interior, which can be seen, e.g., in Fig. 3(g) where the sharpest phase jump now occurs at the secondary DSW oscillation, again consistent with theory.[1,14] In contrast to the DSW, the phase profile of the RW remains relatively unchanged with increased $P_2/P_1$, indicating that the concomitant oscillations are weak and essentially linear. As $P_2/P_1$ is increased, the DSW amplitude relative to the RW amplitude is reduced, also expected. Therefore, the amplitude of the jump is important for DSW development, deriving from the fact that at the jump, both $V_g$ and $D$ vary due to a nonlinearity-induced dispersion shift.

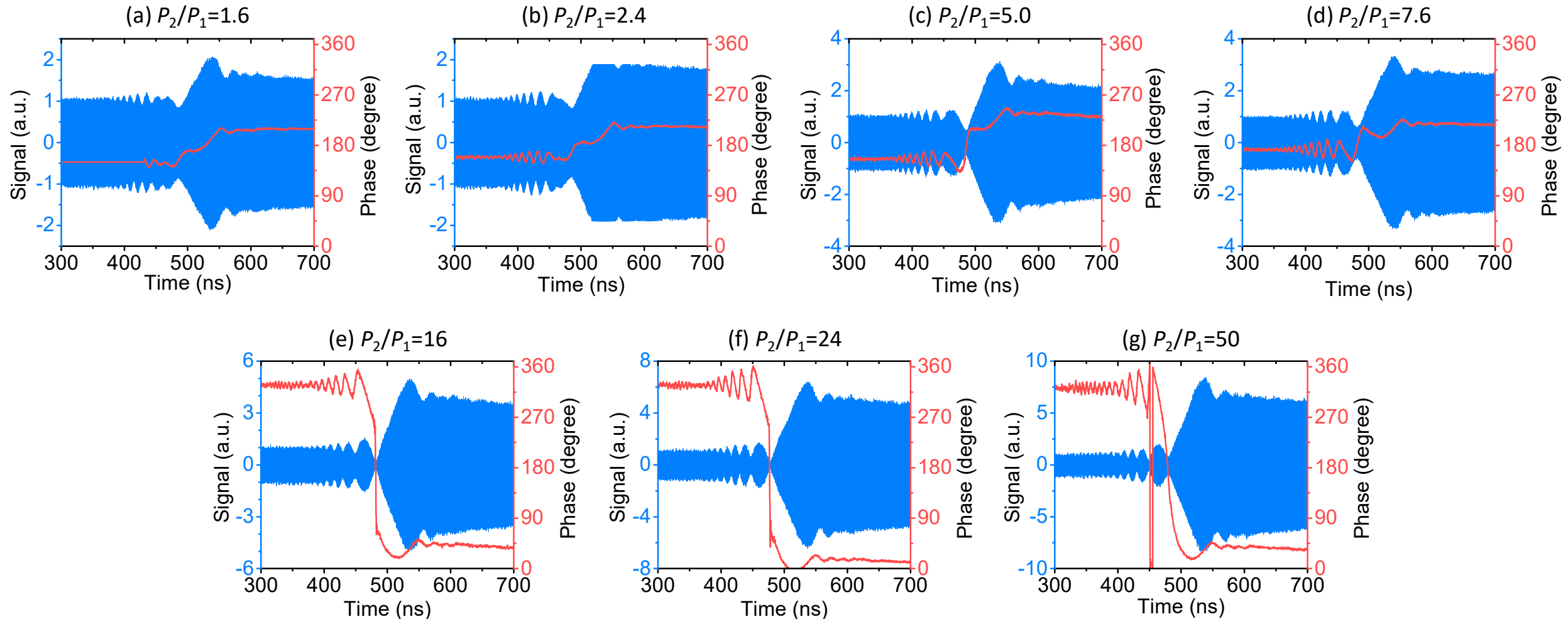

**Fig. 3. Dependence of DSW formation on $P_2/P_1$.** In each diagram, the signal is shown in blue while the phase is shown in red; $P_2/P_1$ is indicated on the top. For all the measurements, $P_2$ was fixed at 3.47 mW.

Figure 4 shows data measured when $P_1$ and $P_2$ were increased but $P_2/P_1$ was kept constant at ~24. The low-power case in Fig. 4(a) lacks the large-amplitude modulation and phase coherence typical of DSWs, while the other cases in Figs. 4(b)-4(e) exhibit nonlinear effects such as increasing amplitude modulations, large phase gradients, and vacuum points. This indicates that appropriate nonlinearity is a pre-requisite for DSW formation. Figures 4(d) and 4(e) exhibit complex dynamics that deviate from the NLS predictions, likely due to higher-order nonlinear processes.

Figure 5 presents data measured at different positions ($x$) along the YIG strip, revealing DSW development. In contrast to those presented in Figs. 1-4, these data were obtained using an inductive probe,[28,29] not a microstrip. For maximum resolution, the figure limits the presentation to the modulations associated with the DSW, neglecting the accompanying RW. The data indicate how an initial step develops into a DSW, revealing the intrinsically expanding nature of DSWs. The fact that DSW formation requires a sufficient propagation length is because nonlinearity requires certain propagation time to develop.[30,31] Different from the black solitons in Figs. 2-4, the dark solitons in the DSWs in Fig. 5 do not show nearly zero intensity at their centers. This is mainly because the propagation distance for the data in Fig. 5 is much shorter than that for the data in Fig. 2-4, while self-cavitation requires a sufficient propagation distance as demonstrated by the simulations. Note that the inductive probe has a lower sensitivity than the microstrip transducer and therefore yields noisy signals when $x \geq 20$ mm.

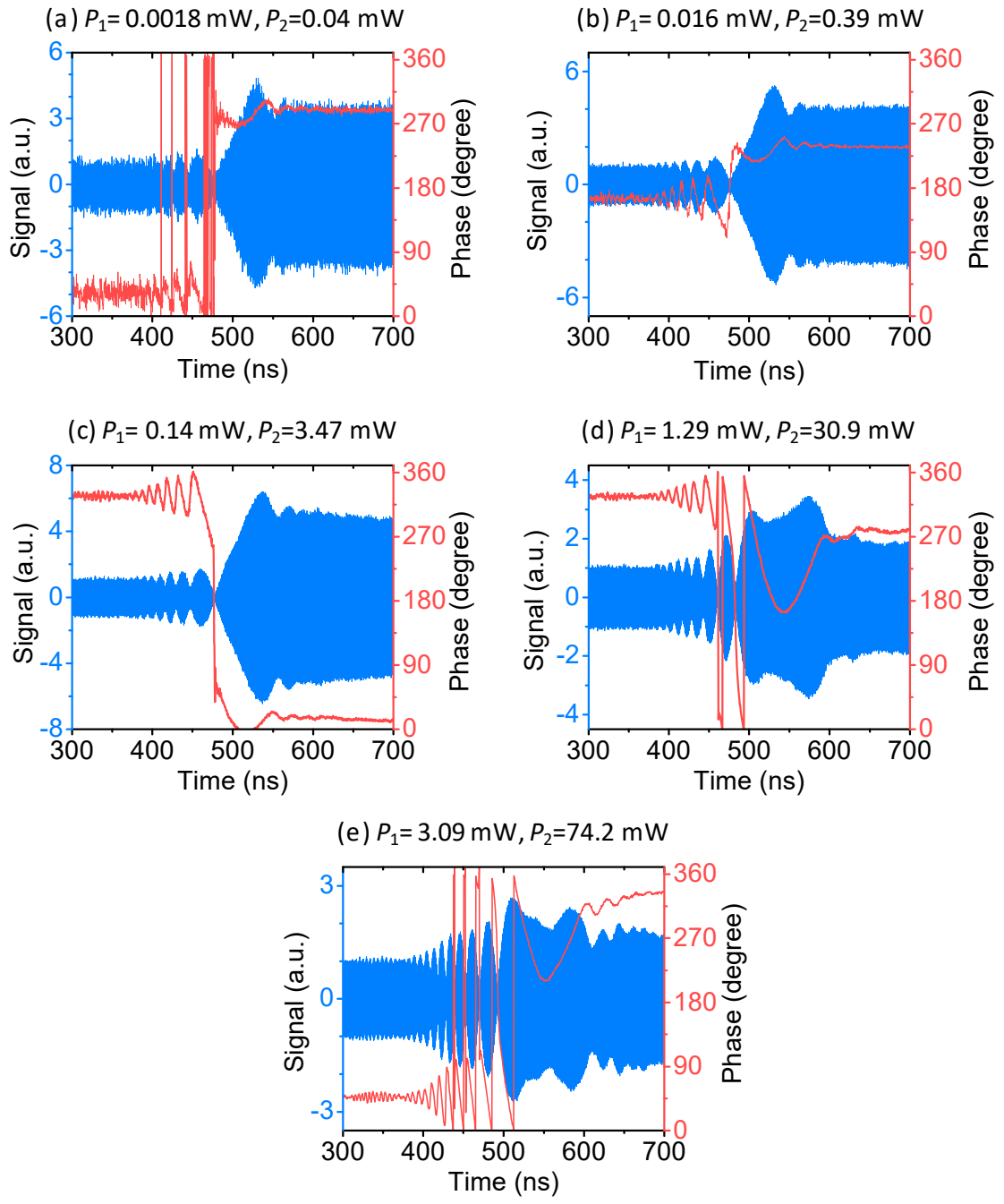

**Fig. 4. Dependence of DSW formation on $P_1$ and $P_2$.** In each diagram, the signal is shown in blue while the phase is shown in red; the input power levels are indicated on the top. For all the measurements, $P_2/P_1$ was kept constant at 24.

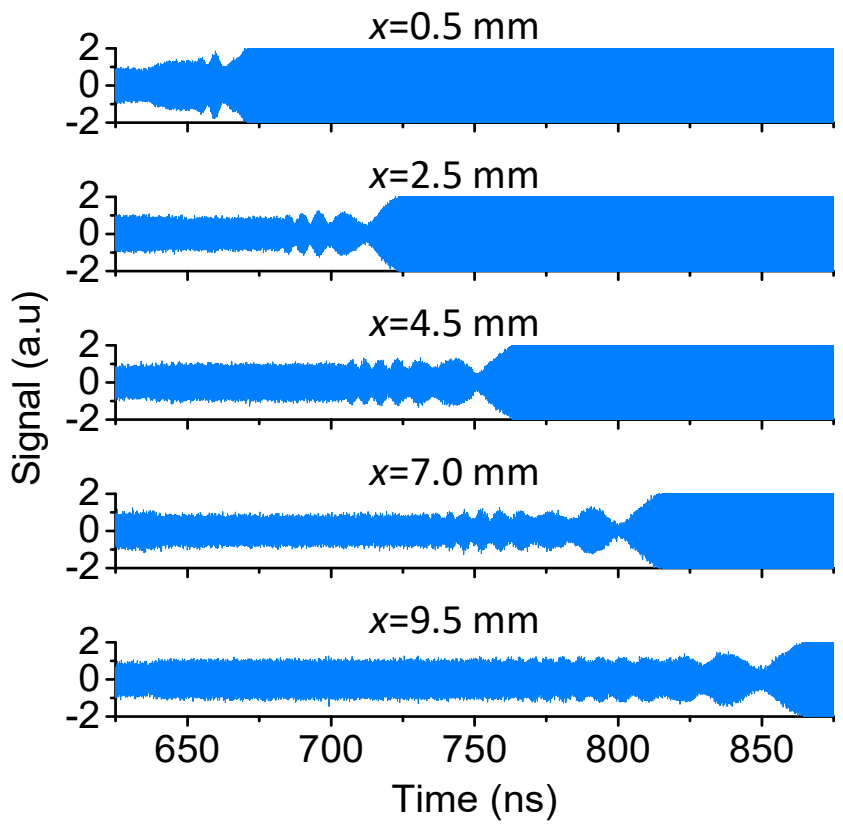

**Fig. 5. Spatial development of DSW.** Each diagram shows the spin-wave signal measured by an inductive probe at a distance of *x* away from the excitation transducer. The measurements were conducted at $P_1$=0.14 mW and $P_2$=3.47 mW.

The experimental observations are reproduced by simulations using the NLS equation with damping. Simulations utilized the parameters associated with the experiments except for an increased initial wave amplitude. The simulation details are given in the Supplementary, while the main results are featured in Fig. 6. The signals in Fig. 6(a) demonstrate the spatial development of the DSW, which is consistent with the data in Fig. 5. The rapid transition time (3 ns) for the initial step results in a Gibbs-type phenomenon[32] with two essentially linear wavepackets at short distances (2.1 mm). At 10.4 mm, nonlinearity enhances the lower wavepacket resulting in a DSW, while the amplitude of the upper wavepacket rarefies into a RW. Consequently, one can trace the weak RW oscillations observed to the rapid step transition. At the output transducer (20.8 mm), the DSW exhibits a vacuum point, which at 101.9 mm has migrated into the interior of the DSW. As shown in Fig. 6(a), the primary role of the damping is to reduce the amplitude by about 20% during the course of propagation. Although not shown in Fig. 6, if a slower step is used, the RW oscillation amplitude is diminished and DSW development takes longer. This suggests that the experimental observations occur in a transient regime, prior to the long-time regime where the solution to the step problem reaches a fully developed state.[14] The fact that the experiment does not clearly show the predicted intermediate state results from a relatively short evolution distance as well as higher-order nonlinearity[33,34] and nonlinear damping[34,35,36] that are neglected in the simulations. Figure 6(b) presents the amplitude and phase results at *x*=20.8 mm, while Fig. 6(c) presents the same data but over a narrower time scale. In Fig. 6(b), the DSW is observed followed by an intermediate transition state. The RW follows with decreasing oscillations that result from the sharp Riemann problem. One can see a remarkable similarity between the results in Figs. 6(b) and 6(c) and Figs. 2(b) and (c), supporting the interpretation of the experimental observations.

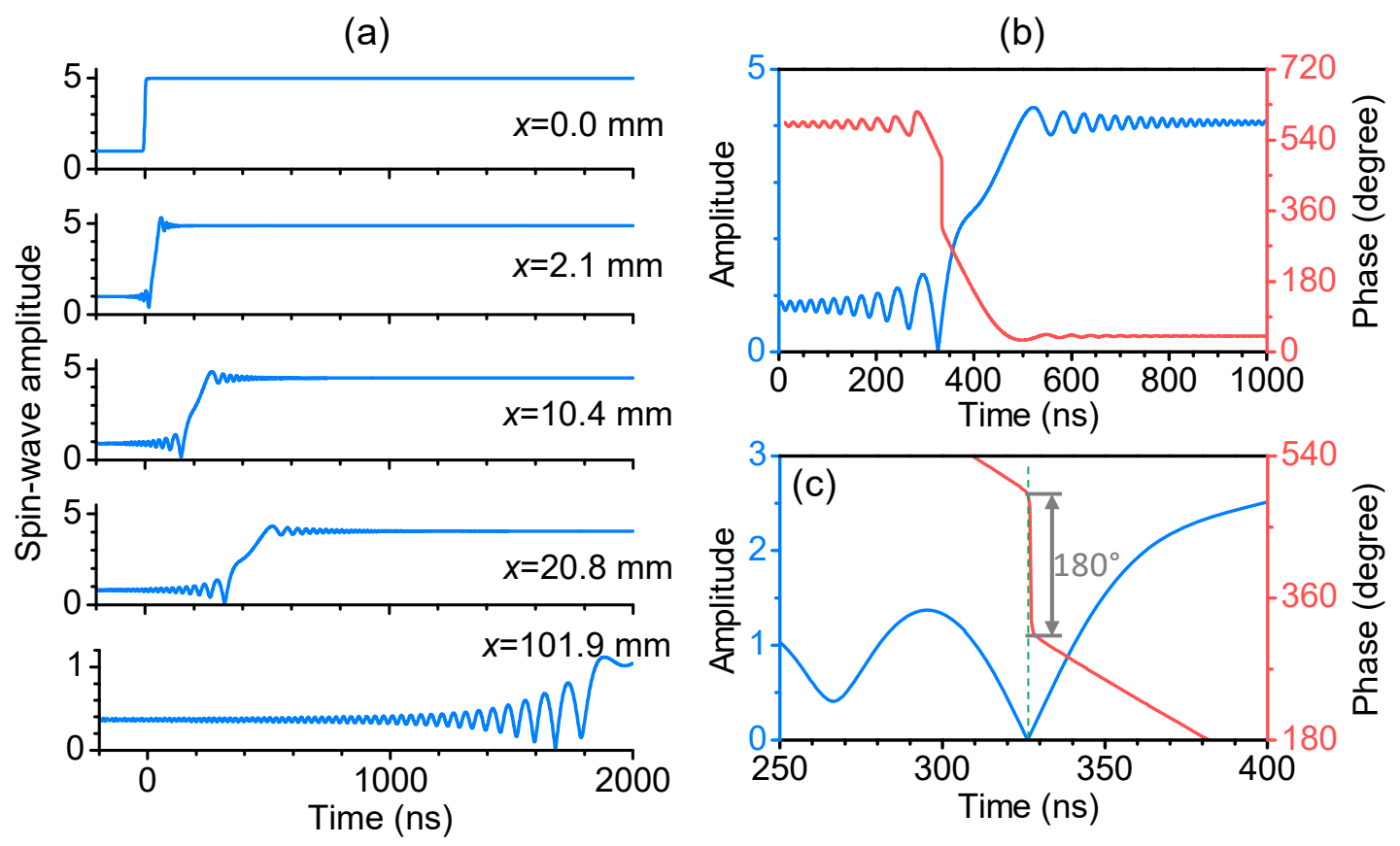

**Fig. 6. Simulation results.** (a) Signals measured at different positions. (b) Amplitude (blue, left axis) and phase (red, right axis) profiles of the signal measured at *x*=20.8 mm. (c) The same data as in (b) in a narrower time window.

In summary, this work demonstrates self-cavitating envelope DSWs for surface spin waves. The DSW consists of a train of dark soliton-like dips with depths increasing from front to back, terminated by a black soliton. DSW formation is sensitive to the characteristics of the initial spin-wave step pulse. A sufficient propagation distance is required for the initial step to evolve into a DSW. Future studies include the exploration of the entire NLS phase diagram for the Riemann problem, DSW and RW interactions, and interaction of DSWs with solitons.

Acknowledgement: The work at Colorado State University was supported in part by the U. S. National Science Foundation under Awards DMR-1407962 and EFRI-1641989; the U. S. Army Research Office under Award W911NF-14-1-0501; the SHINES, an Energy Frontier Research Center funded by the U.S. Department of Energy, Office of Science, Basic Energy Sciences under Award SC0012670; and the C-SPIN, one of the SRC STARnet Centers sponsored by MARCO and DARPA. The work at University of Colorado was supported in part by NSF CAREER DMS-1255422.

*Corresponding author. E-mail: mwu@colostate.edu

---

[1] G. A. El and M. A. Hoefer, Physica D **333**, 11 (2016).

[2] T. Talipova, E. Pelinovsky, O. Kurkina, and A. Kurkin, Hindawi Publishing Corporation, Shock and Vibration Volume, Article ID 875619, **7** (2015).

[34] J. Q. Anderson, R. A. Ryan, M. Wu, and L. Carr, New J. Phys. **16**, 023025 (2014).

[35] M. M. Scott, C. E. Patton, M. P. Kostylev, and B. A. Kalinikos, J. Appl. Phys. **95**, 6294 (2004).

[36] Z. Wang, M. Cherkasskii, B. A. Kalinikos, and M. Wu, Phys. Rev. B **91**, 174418 (2015).